\documentclass[final,1p,times]{elsarticle} 

\usepackage{graphicx}
\usepackage{amssymb} 
\usepackage{amsthm} 
\usepackage{lineno}

\journal{Nuclear Physics A} 

\begin{document}

\begin{frontmatter} 

\title{Theoretical overview of jet quenching}

\author{Jos\'e Guilherme Milhano}
\address{CENTRA, Instituto Superior T\'ecnico, Universidade T\'ecnica de Lisboa, \\ Av. Rovisco Pais 1, P-1049-001 Lisboa, Portugal}
\address{Physics Department, Theory Unit, CERN, CH-1211 Gen\`eve 23, Switzerland}

\begin{abstract} 
In this brief write-up, I overview recent developments on the theoretical description of jet quenching.
\end{abstract} 

\end{frontmatter} 


\section{Introduction}

Jets have since long been regarded as a privileged tool to study the properties of the hot, dense and coloured matter -- the medium -- created in ultra-relativistic heavy ion collisions.
The viability of jets as medium probes relies: (i) on the experimental ability to measure modifications of jet observables -- both for reconstructed jets and for their high-$p_t$ hadronic content -- relatively to a baseline established when no medium is present, what is generally referred to as \textit{jet quenching};  and (ii) on the  theoretical understanding of the relation between properties of the observable collimated spray of hadrons (the jet) to specific modifications imprinted by the medium on its development -- i.e., on the QCD showering of the hard (high-$p_t$) initial parton and subsequent hadronization of the fragments. 

The observation of a strong modification of all single inclusive hadronic spectra at RHIC \cite{Adcox:2004mh,Adams:2005dq}, later confirmed for transverse momenta up to $p_t\sim$ 100 GeV at the LHC \cite{Aamodt:2010jd,CMS:2012aa},  asserted jet quenching as an undeniable experimental reality. 
The extended kinematic reach and detection capabilities afforded by the LHC allowed for the, rather non-trivial \cite{Cacciari:2010te}, full reconstruction of jets in the large and fluctuating background of heavy ion collisions to be successfully carried out and for the observation of  significant modifications \cite{Aad:2010bu,Chatrchyan:2011sx,Chatrchyan:2012nia,Chatrchyan:2012gt} of their properties up to energies in excess of 300 GeV.
The observed dependencies on collision centrality and azimuthal orientation (which provide information on the path-length of in-medium propagation), on kinematical variables (collision centre of mass energy and transverse momentum of the hard process), and importantly the absence of modifications for high-$p_t$ prompt photons, $Z^0$ and $W$, motivate a dynamical picture of jet quenching in which hard partons in heavy ion collisions are produced at standard perturbative rates, but lose energy and branch differently as a result of their interaction with the medium through which they propagate. Recent theoretical developments have been targeted towards extending the descriptions of parton energy loss which, for the most part, were concerned with single hadron (and hadron correlation) observables to the far richer context of reconstructed jets. 

Our present experimental and theoretical understanding of jet quenching has shaped a clear pathway through which jet observables can be used to determine medium properties. In brief, this on-going programme encompasses two essential facets.
First, the establishment of quenched (modified) jets as medium probes, tantamount to a full theoretical account of QCD branching in the presence of a generic medium and a clear identification of the sensitivity of specific observables (or combinations thereof) to particular medium effects. Second, the embedding of such a well defined probe in a realistic account of the medium will allow for the meaningful probing of medium properties.

In this overview, I give a rather personal assessment of the state of progress in the fulfilment of the above programme, focusing on recent developments on the description of the (perturbative) QCD branching, its interface with hadronization, on efforts to identify the relevant dynamical ingredients through comparison with data, and on the implementation of realistic medium descriptions in jet quenching models. Mass effects (the case of heavy quarks) and progress in strongly coupled descriptions are addressed elsewhere in this volume (respectively in \cite{will} and \cite{yee}).  An outline of outstanding challenges closes the discussion.

\section{Defining the probe}

\subsection{Jets in vacumm: a baseline and template}

The excellent theoretical understanding of jet production in vacuum provides both a reliable baseline and a template onto which modifications due to the presence of a QCD medium can be implemented.
A central ingredient of vacuum jet physics is the independence of the jet development (QCD branching and hadronization) on the details of the initial state. In others words, the only relevant initial state information is the probability to find suitable hard partons in the proton wave functions (given by the PDFs) to undergo a perturbatively calculable hard scattering yielding the pair of back to back partons from which the observable jets develop.

QCD branching in vacuum is very well understood in perturbation theory and faithfully implemented in Monte Carlo event generators. Importantly, quantum interference between successive splittings dictate that the shower follows an angular ordered pattern with decreasing emission angles. 

Although not presently understood from first principles, hadronization is effectively described as to be compatible with experimental data. 
In event generators, the partonic fragments are grouped into colour neutral structures  (Lund strings \cite{Sjostrand:2006za}, clusters \cite{Webber:1983if}) which dynamically decay into the final state hadrons. Alternatively, 
fragmentation functions -- probability distributions for a specific final state (jet, hadron class/species, ...) of given momentum fraction to result from the branching and subsequent hadronization of fragments of an initial parton of momentum $p_t$ -- can be defined. They are universal (factorizable) objects of known scale evolution and can thus be constrained via global fits in a manner analogous to the determination of the PDFs. 

\subsection{Jets in heavy ion collisions}

\begin{figure}[tp]
\begin{center}
\includegraphics[width=1.0\textwidth]{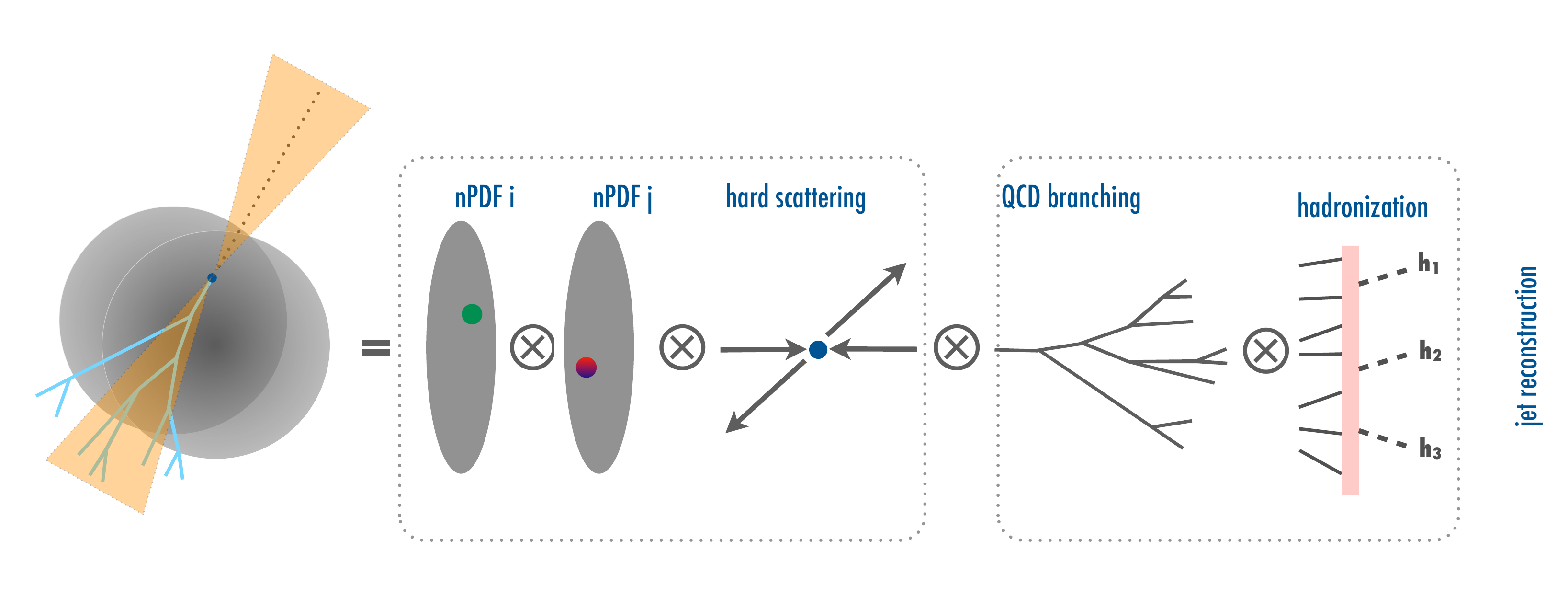}
\end{center}
\caption{The successive stages of jet production in the presence of a medium.}
\label{fig1qm}
\end{figure}
The various stages of jet production in heavy ion collisions are schematically shown in Fig.~\ref{fig1qm}. In this case the probability to find suitable hard partons in the nuclear wave functions is given by parton distribution functions where nuclear effects have been accounted for 
(nPDFs). In the kinematical domain relevant for jet production,  nPDFs are well constrained, and will become more so once results from the forthcoming proton-nucleus LHC run (and from the recent very successful pilot run) are included in the current global fits. 
The hard scattering of these partons occurs localized within a point-line scale ($\sim1/E_t$) and is thus oblivious to its surroundings.  As in the vacuum case, the hard partonic cross section can be computed to arbitrary order in perturbative QCD. 
Taking as a working hypothesis that the factorizable structure encountered in the vacuum case holds also for nuclear collisions, it results that the factorized initial state, shown schematically in the first box of Fig.~\ref{fig1qm}, is insensitive to the produced medium. Although this assertion has been recently challenged \cite{Armesto:2012qa} in that interference between initial and final state radiation was shown to lead to modifications of the angular distribution of the medium induced gluon spectrum, its validity is widely assumed in jet quenching studies which thus attribute the origin of modifications of jet observables solely to effects imprinted by the medium on the QCD branching of the hard parton.  

Prior to the formation of the hot, dense and coloured medium, which occurs on a timescale $\tau_{med}\sim 0.1$ fm, the skeleton properties of the jet are defined by vacuum-like hard branchings.  Effects of the Glasma, the pre-medium coherent state of matter present at this early times, on the gluon radiation rate have been found \cite{Aurenche:2012qk} to be much smaller that those resulting from the subsequent propagation through the medium.  

The jet partonic components traverse a typical medium pathlength of a few fermi. During this time, jet-medium interaction proceeds thorough the exchange of both energy-momentum and colour.
Transverse (with respect to the direction defined by the original hard parton) momentum $k_t$ transfer leads to the modification of the typical radiation formation time, promoting the early emission of soft and semi-soft gluons, and thus induces additional (to that occurring the vacuum) gluon radiation. Further, all jet components accumulate, through multiple exchanges,  transverse momentum ($k_t$ broadening). Parton energy and momentum can be lost to the medium (elastic energy loss). Colour exchanges result generically in the modification of the colour correlations among the partonic fragments and consequently in a disturbance of the coherence properties between successive splittings. 

A significant part  \cite{CasalderreySolana:2011gx} of the branching process, down to the hadronization scale, occurs after escape from the medium. While hadronization happens in vacuum, medium induced modifications of the jet colour structure give rise to a hadronizing system which is, in general, different from that of a vacuum jet.

Finally, the observable jet is defined by a set of rules on how to group the hadronic fragments (jet algorithm) for given defining parameters (e.g., the jet radius) and, importantly, by a procedure allowing for its isolation from the large and fluctuating underlying event (background subtraction).

\subsection{Recent progress I: relaxing approximations}

The limitations of pre-LHC formulations of jet quenching were clearly exposed in the detailed comparison \cite{Armesto:2011ht} of the existing, perturbative QCD based, implementations of medium induced radiation. The four standard formalisms for single gluon emission -- BDMPS-Z, GLV, AMY and Higher-Twist (for references see \cite{Armesto:2011ht}) -- differ on details of the medium modelling and on some kinetic assumptions (although most are shared). All implement multiple gluon emission as ad-hoc iterations -- a Poissonian ansatz for BDMPS-Z and GLV, rate equations for AMY, and a medium modified DGLAP for Higher-Twist Ð of the single gluon kernel.  All have been implemented at the Monte Carlo level (for a review see \cite{Zapp:2011kj}). The formalisms were compared in a well defined common setting -- a brick of static medium of constant density -- and large discrepancies were found.  A key observation of this study is that the, phenomenologicaly necessary, extension of the formalisms beyond their strict applicability domains is, for the most part, the source of the discrepancies.

Important steps have been taken recently to relax some of the assumptions underlying these calculations of single gluon emission. The energy of the radiated gluon is assumed, in all formalisms but AMY, to be much smaller than that of the emitter ($x=\omega/E\ll1$) but the spectrum is computed for all allowed phase space with violation of energy-momentum avoided by explicit cut-offs. The large-$x$ limit, energy of the gluon of the order of that of the emitter, was computed in the path-integral formalism (BDMPS-Z) in \cite{Apolinario:2012vy}. The calculation was complemented by an explicit evaluation in the multiple soft scattering approximation and by an interpolating ansatz generalizing to intermediate x values. 

In a far more ambitious effort, single gluon emission for arbitrary $x$ has been computed in the context of Soft Collinear Effective Theory \cite{D'Eramo:2010xk,Ovanesyan:2011xy,Ovanesyan:2011kn}. This promising powerful framework, non-trivially ported from its high energy genesis to the study jet quenching  in \cite{Idilbi:2008vm}, relies on the natural scale hierarchy of the problem $\textsf{\small{[hard scale]}}\sim\sqrt{s}\sim\lambda^0 \gg \textsf{\small{[jet scale]}}\sim p_t\sim\lambda^1\gg \textsf{\small{[soft radiation]}}\sim\lambda^2$ to define an effective field theory where collinear modes $p_c \sim(\lambda^0, \lambda^2,\lambda)$, soft modes $p_s \sim(\lambda^2, \lambda^2,\lambda^2)$, and Glauber modes $q \sim(\lambda^2, \lambda^2,\lambda)$ -- describing the jet-medium interaction --  are the relevant degrees of freedom. This framework allows for the joint description of elastic and inelastic energy loss, and thus ultimately for the consistent account of medium recoil effects. The elimination of the prevailing artificial elastic-inelastic distinction has been also addressed, in a different formalism, in \cite{Zapp:2011ek}.

\subsection{Recent progress II: QCD (de)coherence}

A bona fide description of multiple gluon emissions, essential for the full description of in-medium QCD branching, requires the understanding of the interference pattern resulting from multiple emitters. Using a $q\bar{q}$ antenna (that is, for emission much softer than both emitters) as a suitable controllable theoretical laboratory a major effort -- initiated in \cite{MehtarTani:2010ma} and continued through an extensive list of works (most recently in \cite{MehtarTani:2012cy} where references to the remaining works by the same authors can be found) and discussed independently in \cite{CasalderreySolana:2011rz} -- to unveil the effect of the medium on QCD coherence is under way.  
In the presence of the medium, the $q$ and $\bar{q}$ colour coherence survives with probability 
\begin{equation} 
	\Delta_{med} = 1 - \exp\bigg(-\frac{1}{12}\hat{q}\theta_{q\bar{q}}^2 t^3\bigg)\, ,
\end{equation}
such that decoherence sets in a timescale 
\begin{equation} 
	\tau_{d} = {\bigg(\frac{1}{\hat{q}\theta_{q\bar{q}}^2}\bigg)}^{1/3}\, .
	\label{eq:dectime}
\end{equation}
Here, $\theta_{q\bar{q}}$ is the antenna aperture and $\hat{q}$ is the typical transverse momentum acquired by unit pathlength. 
Clearly, total decoherence will set in for pathlengths larger than the decoherence time $\tau_{d}$. 
Colour decoherence opens up phase space for emission, leading in particular to large angle radiation. Its effect on the angular distribution of radiation is particularly transparent in the limit of soft radiation $\omega\rightarrow 0$. For an antenna in an overall colour neutral configuration (arising from the decay of a virtual photon), the radiated spectrum reads
\begin{equation} 
	dN^{tot} = \frac{\alpha_s C_F}{\pi} \frac{d\omega}{\omega}\frac{\sin\theta\, d\theta}{1-\cos\theta} \,[\Theta (\cos\theta - \cos\theta_{q\bar{q}}) -\Delta_{med}\Theta (\cos\theta_{q\bar{q}}-\cos\theta)]\, .
\end{equation}
In the coherent limit $\Delta_{med}\rightarrow 0$ the radiation is as in vacuum, limited in angle from above by the antenna aperture (angular ordered). Conversely, for full decoherence $\Delta_{med}\rightarrow 1$ only medium induced radiation is present  and only emission for angles larger than the antenna opening is allowed (the so called anti-angular ordering). 

As interference effects are parametrically suppressed (\ref{eq:dectime}) by $\tau_d/L$ their relevance is restricted, in a multiple emission scenario, for emissions that take place during the formation time of the previously radiated gluon. Thus, in the limit of short formation times the branching process can be iterated probabilistically as a succession of independent splittings and written as a master equation for a generating functional \cite{saclay}.

\subsection{Recent progress III: large broadening effects}

Transverse momentum broadening is an intrinsic part of medium induced gluon radiation. During its formation time $\tau_f \sim \sqrt{\omega/\hat{q}}$ the gluon acquires transverse momentum $k_f^2 \sim \sqrt{\hat{q}\omega}$. This is followed by the classical broadening (Brownian motion) of the produced gluon through its propagation across the remaining medium pathlength. The transverse momentum collected during this later stage $\sim \hat{q}L$ is independent of the gluon's energy. Hence, it can lead, for soft gluons, to very large broadening effects which  lie beyond the quasi-eikonal framework of standard medium induced radiation calculations.  Combined with the observation that soft gluons are formed very early in the medium,  this provides for an efficient dynamical mechanism for the transport of soft partons (all of those with energies $\omega \leq \sqrt{\hat{q}L}$) to large angles without significant disturbance of the direction of propagation of the jet.
This mechanism \cite{CasalderreySolana:2010eh,Qin:2010mn,CasalderreySolana:2011rq} -- jet collimation -- provides a natural and consistent description of the observed dijet asymmetries \cite{Aad:2010bu,Chatrchyan:2011sx,Chatrchyan:2012nia} (and no-modification of their azimuthal correlation) and for of the excess of soft radiation outside the jet cone \cite{Chatrchyan:2011sx}.

\subsection{Recent progress IV: interplay between branching and hadronization}

\begin{figure}[tp]
\begin{center}
\includegraphics[width=0.45\textwidth]{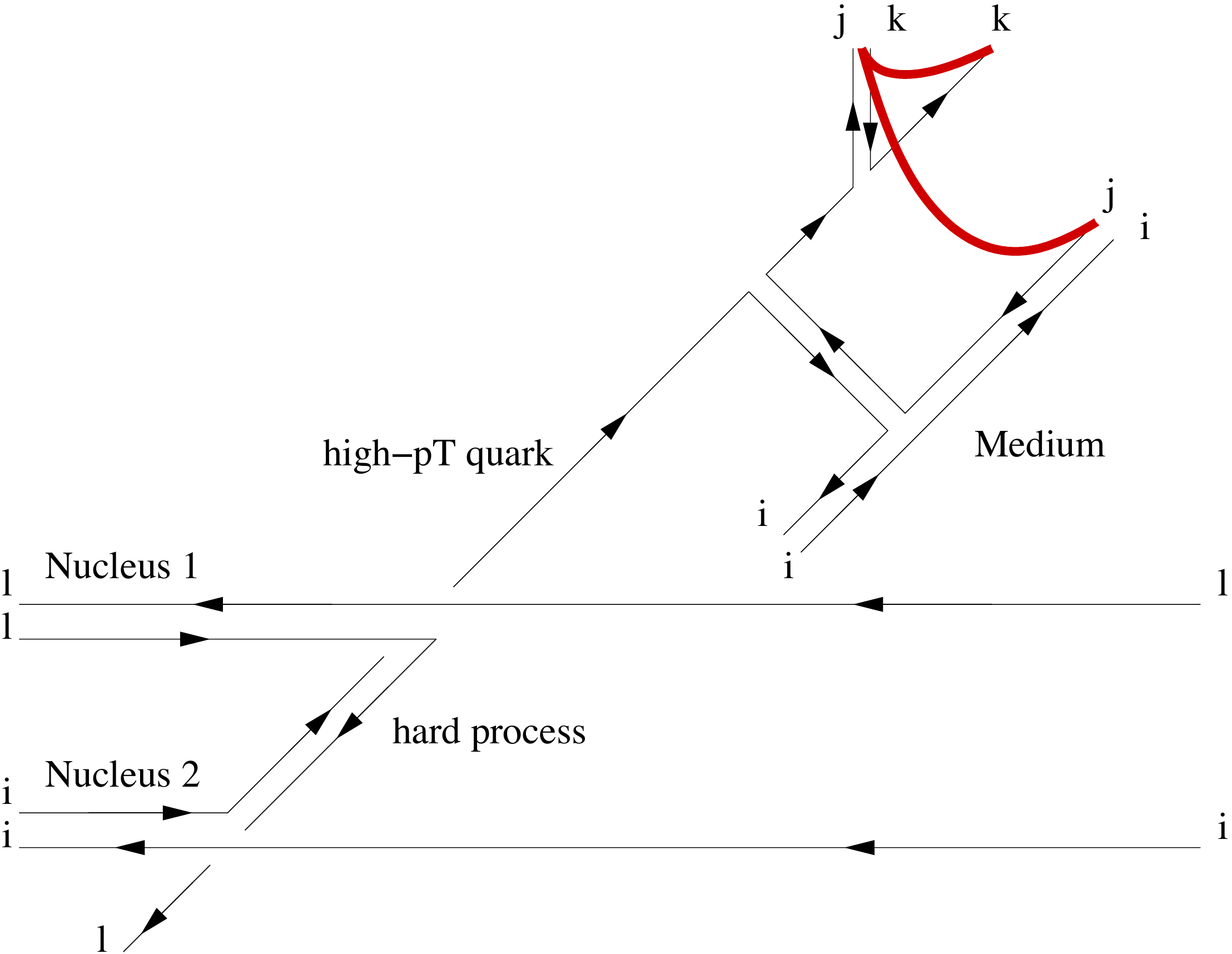}
\qquad
\includegraphics[width=0.45\textwidth]{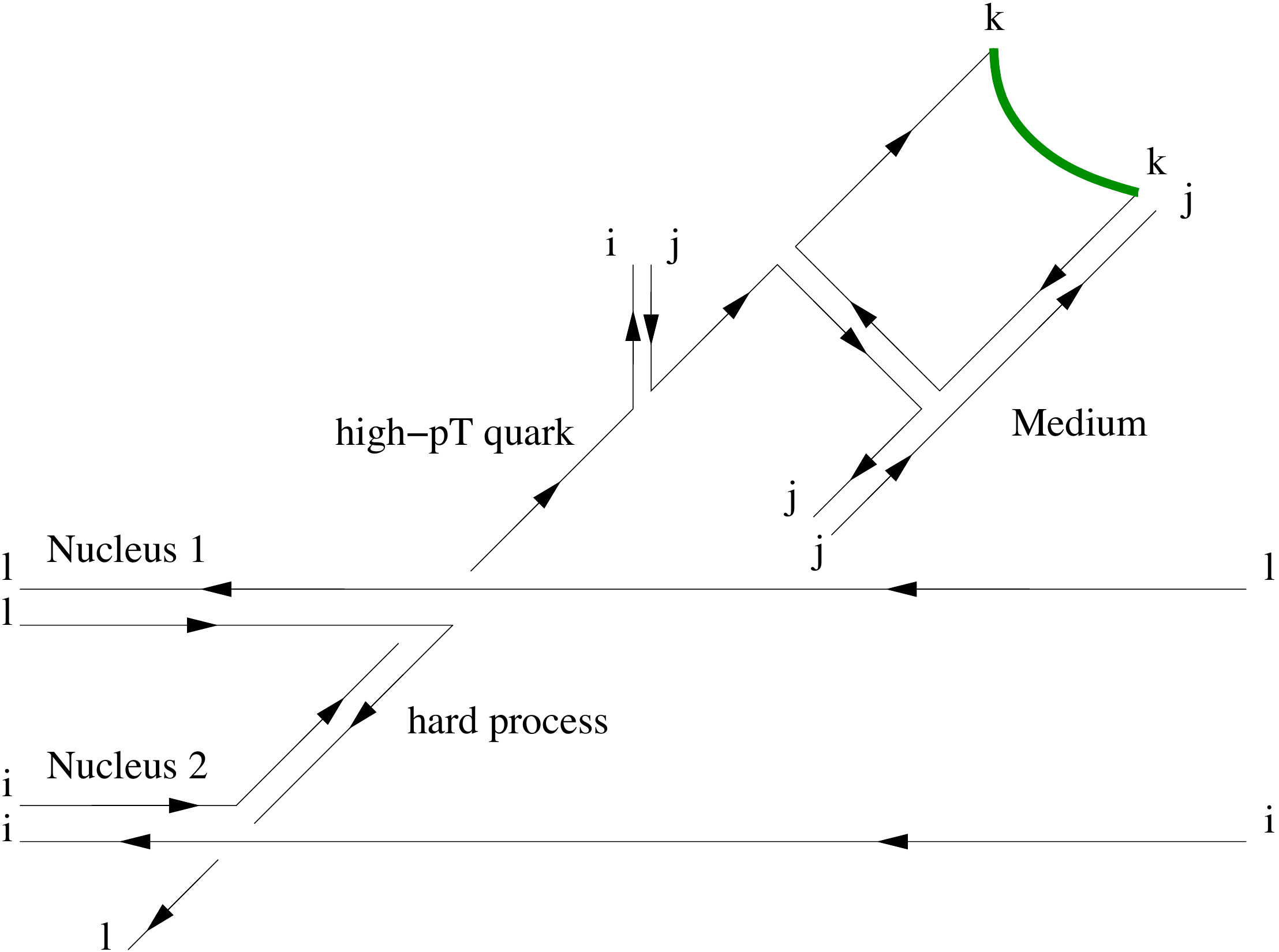}
\end{center}
\caption{Colour configurations (Lund strings) for the leading fragment arising in the colour differential calculation of medium induced radiation. }
\label{fig2}
\end{figure}

The modification of the colour connections among the partonic components of a jet, the result of colour exchanges with medium, is not accounted for in standard calculations of medium induced radiation. The first colour differential calculation \cite{Beraudo:2011bh,Beraudo:2012bq} was carried out in the framework of the opacity expansion and in the proof-of-principle case of a single gluon emission in large $N_c$. For the cases in which no medium interaction takes place after the emission of the gluon, Fig.~\ref{fig2} (left panel), the colour properties of the hadronizing system containing the leading fragment is akin to that in the vacuum. In particular, the radiated gluon belongs to the system. However, when interaction with the medium follows the emission of the gluon, a novel colour configuration arises,  Fig.~\ref{fig2} (right panel), in which the gluon is lost from the leading hadronizing system. Medium interaction of the radiated gluon (not shown) contributes to both colour channels. All in all, medium modified colour structures account for over a half of the radiated spectrum. The modifications survive subsequent, vacuum like branching, and thus will affect hadronization irrespectively of where it happens. When interfaced with realistic hadronization procedures \cite{Sjostrand:2006za,Webber:1983if}, they result in the softening of hadronic spectra with the lost hardness recovered (from the hadronization of sub-leading fragments) in the form of a large multiplicity of soft hadrons. Colour exchanges between jet and medium have been shown \cite{Aurenche:2011rd} to modify the jet hadrochemical content, in particular through anomalous baryon production.

\subsection{Recent progress V: phenomenological consistency}

Ultimately, the virtues and vices of a given theoretical description of jet quenching, and in particular the relative importance of its dynamical components, can only be established by quantitative confrontation with experimental data. It doing so, it is essential that a variety of observables ($R_{AA}$ and $I_{AA}$ for jets and hadrons, dijet asymmetries, jet shapes, jet fragmentation functions, ...) -- for which the specific biases and sensitivity to dynamical ingredients must be understood -- is accounted for. An extensive programme in this direction is underway \cite{Renk:2011qf,Renk:2011aa,Renk:2012cx}. Further essential constraints are afforded by the large RHIC-LHC energy lever arm \cite{Horowitz:2011gd,Betz:2012qq} and by the assessment of the importance of NLO corrections \cite{He:2011pd,Buzzatti:2012dy}

Prerequisites for meaningful comparison of theory and data include the compatible definition of computed/measured jets, the understanding of the response of computables to background effects \cite{Cacciari:2011tm,Apolinario:2012si}, and the folding of detector response effects in the computation of observables for which unfolded measurements are, at present, not available.

\section{Probing the medium}

Two complementary avenues for the embedding of jets in realistic media have been explored. On the one hand, the characteristics of the medium to which jet development is sensitive can be identified and subsequently computed from first principles. To this end, the jet quenching parameter $\hat{q}$ has been computed both in an $SU(2)$ lattice \cite{Majumder:2012sh}. On the other hand, event generators where the best available description of QCD branching and hadronization in the presence of a medium is embedded in a state of the art (3+1) hydrodynamical medium implementation are being developed. Although at present no single such effort is complete (many details of the branching process are not fully understood), the MARTINI Monte Carlo generator \cite{Schenke:2009gb} has established a full framework where future developments can be implemented.

\section{Challenges}

In just over a decade, jet quenching has evolved from being a 'possibility' to a well established experimental reality. An increasingly detailed account of in-medium QCD branching has emerged over the last couple of years. The necessary establishment of quenched jets as medium probes is an on-going programme for which the pathway is clearly outlined: most (if not all) relevant questions have been asked and most are currently in the process of being answered. Notwithstanding this optimistic assessment, 
essential, and challenging, tasks need to be carried out for the successful conclusion of the programme.

The recent recognition of the relevance of (de)coherence effects, of modification of colour flow, of large broadening effects must be followed by their consistent theoretical formulation and implementation in Monte Carlo generators (this appears to be particularly testing for the hadronization of modified colour structures) . It remains unclear whether these effects, which all lead to soft radiation at large angles, are independent or, on the contrary, are simply different manifestations of a common dynamical mechanism.

The insights on the role of decoherence effects gained so far within simplified setups need to be extended to the realistic conditions of an in-medium parton shower. In other words, the interference pattern between successive splittings must be computed beyond the kinematical constraints inherent to the antenna setup.

The wealth of information encoded in the structure of quenched jets remains mostly unexplored. In this context a detailed investigation of sensitivity of jet observables, possibly to be newly defined, to specific medium properties is particularly urgent.

Encouragingly, the necessary tools for the embedding of faithful accounts of QCD branching on a realistic medium are already in an advanced stage of development.

\section*{Acknowledgements}
I wish to thank the QM2012 Organizing, International Advisory and Program Committees for the invitation to overview the theory of jet quenching. My work and conference participation were partly supported by the Funda\c c\~ao para a Ci\^encia e a Tecnologia (Portugal) under project CERN/FP/123596/2011.

\end{document}